\documentclass[peerreview,10pt,letterpaper]{IEEEtran}
\IEEEoverridecommandlockouts
\usepackage{cite}
\usepackage{amsmath,amssymb,amsfonts}
\usepackage{algorithmic}
\usepackage{graphicx}
\usepackage{textcomp}
\usepackage{xcolor}

\newcommand{\cxxcode}[1]{{\texttt{#1}}}
\newcommand{\javacode}[1]{{\texttt{#1}}}

\begin{document}

\title{Experience on Automatically Converting a C++ Monolith to Java EE\\
}

\author{\IEEEauthorblockN{1\textsuperscript{st} Andre Vehreschild}
\IEEEauthorblockA{\textit{Badger Systems GmbH}\\
Köln, Germany, vehre$@$badgersystems.de}
\and
\IEEEauthorblockN{2\textsuperscript{nd} Lexi Pimenidis}
\IEEEauthorblockA{\textit{Badger Systems GmbH}\\
Köln, Germany, lexi$@$badgersystems.de}
}

\maketitle

\begin{abstract}
Converting a large C++ code base (800k lines of code) into Java alone is challenging. Changing the architecture from a monolith into an application adhering to the Java application server standard and to run it on WildFly is a different number. This report describes the experience made during the C++ to Java conversion, the techniques used as well as the way to success of running the Java code on the application server for the first time. The approaches to solve the usual C++ to Java culprits, like multiple inheritance, enum-handling and scoped objects are described. A clang-tool-based software is developed to continuously regenerate the Java, because development on the C++ code base continued.
\end{abstract}


\section{Introduction}\label{sec:introduction}

Switching the programming language of a 800k lines of code C++ artifact to Java poses some challenges. With ongoing development like feature additions and bug removal a manual approach seems too tedious and is prone to errors. The llvm/clang-tool~\cite{LLVM:CGO04} provides a stable and well tested C++ parser and offers an easy to use programming interface. An automatic tool to convert C++ to Java is therefore introduced and the challenges and solutions perceived are reported here.

While this section introduces the task at hand, section~\ref{sec:motivationAndChallenges} details the motivation for the endeavor and describes the obstacles in the path. Section~\ref{sec:theTranspilerInGeneral} outlines the general structure of the tool and some standard techniques put to use. Approaches to solve the challenges posed by the artifact to convert are given in section~\ref{sec:addressingTheChallenges} in detail. At the end, in section~\ref{sec:conclusionAndResults}, the success is shown by the decrease of the error count while the challenges are resolved.

\section{Motivation and Challenges}\label{sec:motivationAndChallenges}

In 2015 a company for logistics and related software decided to unify the base layer of all their software products. The decision was made to provide just one logging system, one user administration and one database management and so on for all the different products all departments of the company sold. It was furthermore decided to switch from C++-98~\cite{ISO:1998:IIP} to Java-8~\cite{Java-8} with the consideration, that Java developers are more plentiful and therefore expected to be easier to hire. Different departments of the company took different approaches to convert their C++ artifacts into Java programs. One department would rewrite completely from scratch, another converts manually using kinds of mechanical turks~\cite{mechturk} and the one reported about here decided to use software to do the conversion by a press of a button. Another hidden agenda was that development on the C++ code base had to continue while the conversion is in progress and a software driven conversion would in an ideal world allow to develop in C++, use the converter and get a Java artifact which would run without any manual intervention. This of course was never accomplished in the real world. Albeit are the remaining manual interventions quite limited.

The task of the authors was to support in the conception and development of the converter, hereafter often called transpiler from \emph{trans}formational com\emph{piler} (see \cite{transpiler} for an interesting history on the topic). The company already had done some research on the C++ to Java conversion task and deemed those available~\cite{tang:cpp2javaconv, mohca}  as too inflexible at the time of evaluation. The company therefore choose to use the clang-tool~\cite{LLVM:CGO04} as a starting point for the analysis of the existing C++ code base.

The artifact to analyze is a standalone server-only application implemented as configurable monolith, i.e., depending on the configuration the same executable acts as information broker, yard controller, print server or transport optimizer. The C++ classes relied only on a small number of libraries that are not part of standard C++. Mostly those enabled accessing hardware, like the serial port for barcode~\cite{barcode} scanners. Furthermore, some classes of the STL~\cite{stl} library are used, where the list of \cxxcode{std::vector<>}, \cxxcode{std::iostream} and \cxxcode{std::map<>} nearly gives the complete set.

Particularly the artifact does not contain any graphical user interface. User interaction is done through proprietary messages send via socket communication~\cite{sockets} encapsulated in a http~\cite{http11} kind of way. Table~\ref{tab:stats} gives some numbers on the classes, files and lines of code.
\begin{table}[btp]
\caption{Artifact's statistics}
\label{tab:stats}
\centering
\begin{tabular}{|l|r|}
\hline
\multicolumn{1}{|c|}{\textbf{Category}} & \textbf{Number (approx.)} \\
\hline
files (*.cpp) & $2\,500$ \\
classes & $3\,000$ \\
lines of code & $800\,000$ \\
\hline
\end{tabular}
\end{table}
The monolith does not use multi-threading, but uses a kind of MPI-like~\cite{mpi} communication protocol between its distinct instances of the monolith.

A database connection to Oracle databases uses DAOs~\cite{dao} generated by a self-written tool based on the database schema language as used by the official Oracle tool~\cite{bryla_loney_2007} with some extensions. The generated DAOs are not to be fed into the transpiler, but the self-written tool is to be adapted to generate Java DAOs instead.

The authors of the monolith are proud of their extensive logging. It not only enables error tracking, but also allows for comprehensively tracking of transport units and their where-about. By using indentation the developers can identify why a certain storage process came to the result at hand and what the consequences of these actions in finalizing the process are. Essentially this is done by creating an object on the (C++) stack, which increases the indentation on its instantiation and consequently decreases it on destruction. This allows to easily track start and end of a transaction even if multiple ones transpire at the same time.

\begin{figure}[btp]
\centering
\includegraphics[scale=0.5]{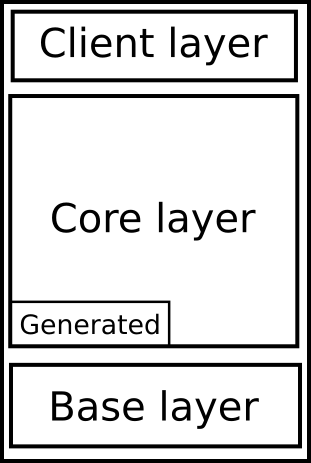}
\caption{Component structure of the monolith}
\label{fig:monolithstruct}
\end{figure}

The monolith to convert from C++ to Java essentially has the structure as depicted in figure~\ref{fig:monolithstruct} where the sizes of the boxes give an idea of their share in the total amount of code. The base layer here supplies the core routines, like file IO, logging, database access and some other. The core layer has the routines doing the work like routing goods, yard management or printing barcodes, package lists and so on. The routines model the base for the client's layer, which is written for each client and installation specifically. The client layer's size varies for different installations depending on the hardware used or other special requirements of the client. The core layer furthermore contains some generated classes, most of them DAOs. The client layer often also has a small amount of generated code, but, because the whole client layer it not converted at the moment, this is ignored here.

The clang-tool's interface was initially used to figure what kind of concepts and idioms are in use in the monolith's core layer. Listener~\cite{design_patterns} classes of the clang-tool were extended to count and report several concepts and idioms, that were deemed problematic in the conversion of the code base. An incomplete list (without any rating) of the concepts and idioms tested for is:

\begin{itemize}
\item multiple inheritance
\item stream in- and output (\cxxcode{std::ostream}, \cxxcode{std::istream})
\item constructors which do more than delegate to their base class
\item destructors with code
\item enum-integer-equivalency (assigning an enum to an int and vice versa)
\item throwing exceptions
\item handling of exceptions
\end{itemize}

Those concepts and idioms that were not found in the code base of the core layer are not explicitly named here. An example for such an idiom is C-style IO, which was only present in the base layer which was not scope of the analysis.

While some of the concepts and idioms named above are a challenge to convert, because Java does not support it, like multiple inheritance, destructors or enum-integer-equivalency, are others only challenging when the targeted software architecture is known. The company had decided to not only convert from C++ to Java, but also to use a Java application server. This means that the architecture of the software has to be changed during conversion as well. This motivates rating the other concepts and idioms as challenging. The IO in a multi-threaded world needs to be protected against concurrent access. Exception handling needs to be put in place correctly. In the past an exception usually terminated the program, because error situations were handled using return codes. The constructors doing more than base class delegation where investigated, because the Java application server could reuse a once created object. I.e. code with side effects called by constructors no longer would be executed reliably. Destructors on the other side are simply not part of the Java language specification and therefore any code in a C++ destructor would never be execute even when translated to Java.

The most challenging concept for the conversion was multiple inheritance. Table~\ref{tab:mulinh} gives some numbers on the classes that inherited from one or more base classes. While the classes using only one base class are not a problem, the ones with more than one base class needed a more thorough analysis.
\begin{table}[btp]
\caption{Classes using one or more base class}
\label{tab:mulinh}
\centering
\begin{tabular}{|l|r|}
\hline
\multicolumn{1}{|c|}{\textbf{Inheritance}} & \textbf{Number (approx.)} \\
\hline
one base class & $1\,200$ \\
more than one base class & $400$ \\
\hspace{1em}split up into: & \\
\hline
\hspace{1em}one DAO, one other & $200$ \\
\hspace{1em}one Chain, one other & $150$ \\
\hspace{1em}one Chain, one DAO & $30$ \\
\hspace{1em}multiple, including DAO and Chain & $20$ \\
\hline
\end{tabular}
\end{table}
Of the about $400$ classes inheriting from multiple base classes, $380$ inherited from two base classes only, where one was either a DAO-class or a Chain-class (Chain of Command~\cite{design_patterns}). But also about $30$ classes inherited from a DAO and a Chain class. These cases could be easily handled, which is shown later. More challenge posed the $20$ remaining classes, where one extraordinary class had five base classes.

Besides these challenges the company also decided to make use of the Java package mechanism. C++ namespaces were scarcely used in the core module. Most of the time class names encoded a notion of package by using two or three letter abbreviations of the German term. 'Te', for example, denotes the German word for transport unit (a box, crate or pallet), while 'Wem' is used for incoming goods management (in German: '{\textbf W}aren{\textbf e}ingangs{\textbf m}anagement'). A class registering a transport unit in the incoming goods management therefore is prefixed by 'WemTe...'. These virtual namespaces should be split up into real packages. Fortunately the old abbreviations are reused for the components of the package's names to support developers in finding there way through the converted code once development is no longer done in C++ but in Java.

\section{The Transpiler in General}\label{sec:theTranspilerInGeneral}

\begin{figure}[btp]
\centering
\includegraphics[scale=0.55]{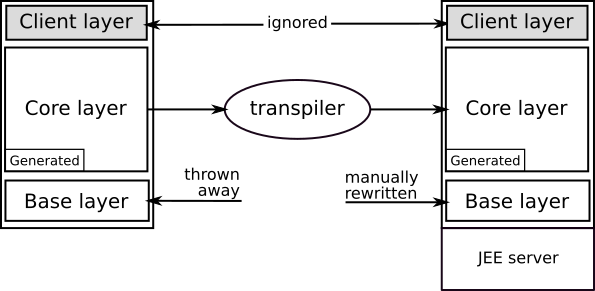}
\caption{General conversion concept}
\label{fig:genconcept}
\end{figure}

The transpiler's purpose in an ideal world is as depicted in figure~\ref{fig:genconcept}. The C++ base layer is thrown away and a new interface compatible version in Java is written. The generators for all generated code are adapted to emit Java code. In the first stage of the project the client layer is omitted. The core layer is able to run without a client layer. The transpiler's executable is then responsible for translating all C++ classes without any user interaction into valid Java-code.

Unfortunately is the world not ideal and the transpiler is only able to convert most input files into valid and correct Java. Nevertheless are in the end about ten classes left, that need some manual tweaking. Those could have been treated especially in the transpiler, but the effort would not have paid. It was decided to implement a checked replace. In the process of converting the whole code base, the expected incorrect Java output was compared to a template (checked) and if those matched the file was replaced by a manually corrected Java file. This was implemented in the build system to be able to automatically convert the C++ code base on each check in into a code repository. Keep in mind, that the development of the C++ artifact was not discontinued. This procedure was put in place to get notified about changes of the problematic classes and adapt the correct Java accordingly.

\begin{figure}[btp]
\centering
\includegraphics[scale=0.57]{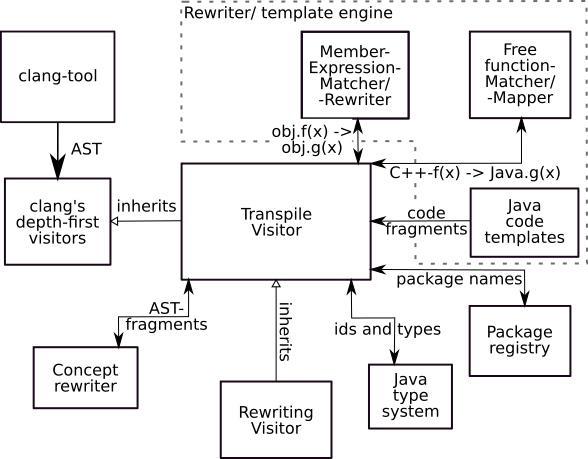}
\caption{Overview of the internal structure of the transpiler with some dataflow}
\label{fig:intstruct}
\end{figure}

The transpiler itself is written in C++, which is motivated by the llvm/clang-framework it's based upon. Around a central transpile visitor~\cite{design_patterns} various infrastructure components are placed as depicted in figure~\ref{fig:intstruct}. The transpiler is feed one C++ file after the other. There is no central data storage to keep track of artifacts already transpiled, because each C++ file needs to be correct C++ as clang understands it, which makes all identifiers in it well defined. The clang-tool front-end ensures that all header files necessary are read and all symbols in the C++ file are resolved. It prepares an AST for further processing. In the logistics software by convention the file name of the implementation (the .cpp-file) always has the same name as the class it implements. To only transpile that class the AST is fast forwarded to the first class-declaration of that name. Forward declarations are especially skipped. Next the transpile visitor is started on that class-declaration. It looks up all member and method declarations in the AST. Usually it is completely sufficient to process the AST upto its end to get the complete class. This held for all classes, i.e., no classes were present that had their implementation distributed to several .cpp-files.

The transpile visitor inherits from several depth-first visitors of clang. The most prominent being \cxxcode{clang::ConstDeclVisitor<>}, \cxxcode{clang::ConstStmtVisitor<>} and \cxxcode{clang::TypeVisitor<>}. The \cxxcode{ConstDeclVisitor} takes care about all declarations (the \cxxcode{Const} in the name denotes, that the AST is unchangeable and not that only constant declarations are processed). The \cxxcode{ConstStmtVisitor} is used to process everything from the simple assignment to expressions and flow control statements. Last but not least is the \cxxcode{TypeVisitor} necessary to process the in-class type declarations and to traverse the whole class type. Furthermore, it does support handling of enums and typedefs.

Although the transpile visitor is the workhorse in the transpiler is there another layer on top, the rewriting visitor. The rewriting visitor extends processing of some expression visit methods, analyses the arguments more thoroughly and emits optimized output. This visitor was intended to do more in-depth analysis and optimization than it is is finally doing. The decision was made, that the Java code should resemble the C++ as much as possible. This reduced the rewriting visitors functionality to essentially only rewrite boolean expressions where one operand was the fixed value \cxxcode{true} or \cxxcode{false}. Those expressions were not as scarce as one would expect. The reason for that is unknown.

Figure~\ref{fig:intstruct} shows in its upper right corner the Rewrite/ template engine. It is composed of three parts. One part is the Java code templates library, which for nearly each C++ construct gives the Java counterpart. Figure~\ref{fig:strtempcast} gives a tiny excerpt of all templates. On the left-hand side of the \verb|:=| an identifier is given which is hard coded into the transpiler. I.e., when the transpiler visitor experiences an explicit cast expression, it figures the \verb|type| to cast to for the \verb|expr|ession to cast. Both symbols are also hard coded into the transpiler visitor. The type/expression to substitute for them are stored in a map. These substitutes are read from the map whenever a symbol in double curly parentheses is found in the text template when emitting it. A symbol in double curly parentheses that can not be found in the map is an error and leads to termination of the conversion. Figure~\ref{fig:strtempcast} shows some examples for standard cast expressions and selected special casts, like the implicit cast of a \javacode{bool} to an \javacode{int} or the other way around from an \javacode{int} to a \javacode{bool}. Note, that Java has only very limited implicit cast mechanisms (most of them are (un-)boxing). I.e., the C++ statement \cxxcode{if (42) something;} in Java has to become \javacode{if (42 != 0) something;} (ignore that the code can be significantly optimized here). Initially the text template approach was chosen to allow the company to easily change the style of the emitted code. Later on the style was defined using a third-party tool, which was applied after the conversion process as part of the build or explicitly in an IDE.

\begin{figure}[btp]
\begin{verbatim}
gen_cast:=(({{type}}) ({{expr}}))

bool2int:=(({{expr}})? 1: 0)

int2bool:=(({{expr}}) != 0)
\end{verbatim}
\caption{Java templates for cast expressions}
\label{fig:strtempcast}
\end{figure}

A smaller part of the rewriter/template engine is the free function matcher/mapper depicted in the upper right corner of figure~\ref{fig:intstruct}. This part takes care about C-style free functions used in C++, like \cxxcode{atoi()} or similar but also custom ones. In Java no free functions exist, therefore those have to be mapped to some static class functions. The mapper is a rather simple table where the first column gives the free function's name. The second column gives the package and class where the Java counterpart is to be found. When the free function from C shall have a different name in Java, then the third and last column gives the new identifier. This mapper is quite simple, it does not respect the number of arguments nor the type of those. It it is rather scarcely used and the limited parametric polymorphism is handled in the Java classes by overloading. Therefore any more elaborate design did not pay.

The third and last part in the rewriter/template engine is the member expression matcher/-rewriter. This part cares about member functions and overloaded operators of standard C++ classes, like \cxxcode{std::string} or \cxxcode{std::map}, but also about some member functions of the C++ base layer that are renamed, moved or likewise replaced in the new Java base layer. The list of member functions to map/rewrite for a C++ class is given by a table. The first column gives the name of the C++ class including its namespace, e.g. \cxxcode{std::string} for the standard string class. For template classes, like \cxxcode{std::map}, the template arguments are given here as well. In case of the map this is resembled by \cxxcode{std::map<KEY, VAL>}. The next column gives the signature of the method to rewrite. The only difference to the standard C++ method definition is, that the return type has to be at the end of the signature definition and not a the beginning. Furthermore, is the class-type omitted. For each overload of a method a separate row needs to be given. Furthermore, default arguments are not supported, but have to be modeled using two rows, one for the call with using the default value and one without. The third column gives the type signature of the resulting Java expression, where the method-name is omitted and replaced by the Java-class-type. I.e., the Java type information of a call of the index method of the C++ string class (\cxxcode{char std::string::at(size\_t)}), is given by \javacode{char String(int)}. This tells the expression rewriter, that the only argument has to be an int. The object the method is applied to has to be of the Java class type \javacode{String} and that the result of the expression is a \javacode{char}. The last column gives the Java-code to use. In this expression \$-symbols denote the placeholders for the arguments. Here \$0 gives the object, while \$1 to \$n give the first to n-th parameter.

\begin{figure*}[!tbp]
\begin{verbatim}
std::string, operator[](unsigned long long) char,\
    char String(int), $0.charAt($1)

std::string, find(std::string, unsigned long long) unsigned long long,\
    int String(String, int), $0.indexOf($1, $2)

std::string, find(std::string) unsigned long long,\
    int String(String), $0.indexOf($1)

std::string, find(char, unsigned long long) unsigned long long, \
    int String(char), $0.indexOf((int)$1, $2)
\end{verbatim}
\caption{Example on member expression rewrites of a few \cxxcode{std::string} operators and functions.}
\label{fig:stringrewrite}
\end{figure*}

In the final implementation the table has some more columns to give some flags for specific handling of the member function and to add necessary package imports. For brevity these have been omitted here, because the added value is insignificant. In figure~\ref{fig:stringrewrite} a few example rows of the expressions for the \cxxcode{std::string} class are given. The lines are broken by a backslash for better readability. The first row gives the rewrite for the \cxxcode{std::string} index expression and its counterpart in Java \javacode{String.charAt(int)}. The remaining rows give an example for the overloaded \cxxcode{find()} method. The first instance shows the find of one string in another giving a starting position. The second instance essentially denotes the same C++-method as the instance before, because the second argument has a default value in C++. Here an extra row has to be specified to neatly map this member call to a Java member call without cluttering the Java-code with dropped in default values. Furthermore, the default value of C++ can possibly be problematic for the Java method (like being out of range). Therefore, the third row gives the find method with just one string argument. This will be mapped to the unary \javacode{indexOf()} method of Java's string class. The last row in the example shows an overload of the find method where the first argument is only a single \cxxcode{char}. This of course is just a very small subset of all definitions of the \cxxcode{std::string} class. In the productive member expression rewrite table there are more than $100$ rows for overloaded operators and member functions of the \cxxcode{std::string} class. In total more than $400$ rows are in the member expression rewriter table giving sufficient definitions to convert the artifact at hand. To specify sufficient many rewrite rules to allow conversion of any arbitrary C++ artifact more than thousand rows are to be expected in the expression rewrite table. Because not every C++ class library function has a compatible counterpart in Java some functions have to be mapped to manually written static Java class methods of some support class.

Another difference between C++ and Java the transpiler had to take care about is the different way of handling an import of external code. While C++ uses the preprocessor employing includes, Java is using imports of packages. Another big difference is that C++ can include in include files and expose the symbols for the sub-includes. The artifact to convert used this in its extreme, where most implementation files included their header and an include-all-header file. Fortunately this was changed in the artifact to include only the headers really necessary for the class file to compile speeding up not only the C++ compile but also the conversion to Java. The company also decided not to map include files to imports, but to associate packages with symbols. I.e., when the expression member rewriter was triggered to rewrite an expression on a STL-class, it also registered the necessary package to import with the package registry. Furthermore, some transpiler visitor methods are coping with type declarations deducing the packages to import from the C++ type name by the two or three letter prefixes. All of these packages, which deemed necessary for the Java class, are registered in the central package registry, which is implemented as a singleton to be available from everywhere. On emitting the final class file, the package registry adds the necessary import statements. This works so well, that in the end no errors are left because of missing imports.

\begin{figure}[btp]
\centering
\includegraphics[scale=1]{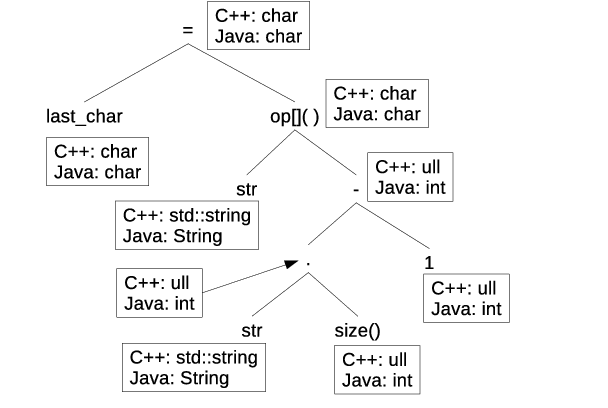}
\caption{C++ types mapped to Java types for the expression \cxxcode{last\_char = str[str.size() - 1]} (\cxxcode{ull} is the typedef for \cxxcode{unsigned long long}).}
\label{fig:cxxjavatypes}
\end{figure}

In the first stages of the transpiler only a C++ type system was present, which was provided for free by the clang-tool. Unfortunately did it not proof very feasible to rely only on the types of the C++ side. An arbitrary example why this did not work as expected is the \cxxcode{std::string::operator[]}. The type of the single argument of this operator in C++ is some typedef which, depending on the C++ standard used, is declared in either the \cxxcode{Allocator} or the \cxxcode{std::allocator\_traits} class, respectively. To simplify the definition of member expression rewrite rules it was decided to resolve this type to its canonical type. For the example at hand this lead to the \cxxcode{unsigned long long} data type on a 64-bit system, i.e., to the intrinsic C++ data type where all type definitions have been followed. In more recent versions of the transpiler a Java type is passed along with the C++ type when traversing the AST. A simple example is depicted in figure~\ref{fig:cxxjavatypes}. Here the simplified AST for the assignment \cxxcode{last\_char = str[str.size() -1]} is shown (nodes without a direct code relationship, like implicit casts, have been removed to prevent clutter). Allow the symbol \cxxcode{str} to be declared as \cxxcode{std::string} and \cxxcode{last\_char} as \cxxcode{char}. When the AST is produced by the clang-tool, only the C++ type information is present. Here the canonical type \cxxcode{unsigned long long} is abbreviated by \cxxcode{ull}, because that type name is really long and would only have bloated the figure without adding content. When traversing the AST depth-first from left to right for nodes on the same depth, the Java type for \cxxcode{last\_char} is retrieved from a symbol registry, that stores any previous declaration of identifiers. The next node to be examined is the \cxxcode{str} in the center of the figure, which also is resolved by a lookup in the symbol registry. The same accounts for the \cxxcode{str} on the bottom left of the AST. With the member-call of \cxxcode{size()} things start to get more interesting: The member expression rewriter matches the size method of a \cxxcode{std::string} object to have \javacode{int} as its resulting Java type, which is propagated by the member call operator (\cxxcode{.}) to the intrinsic subtraction operator (\cxxcode{-}). The minus already knowing its left-hand operator's Java type sends this along when traversing its right-hand operator to indicate that it rather would prefer this type instead of some implicitly converted type when possible. The constant valued node \cxxcode{1} examines the preferred type and if it is compatible rather returns that type instead of allowing for any implicit intermediate type casts. In the end the rather neat Java-code \javacode{last\_char = str.charAt(str.size() - 1)} is emitted instead of the more unreadable mechanically translated expression \javacode{last\_char = str.charAt((int)((long)str.size() - 1L))} if the expected type would not have been taken into account.

The Java type system not only prevents generation of too many casts, but also supports in tracking enums correctly. In C++ enums can be converted into ints or a similar numeric data type implicitly. Converting a numeric value into an enum is also easy in C++. Just a cast is needed to allow assignment of any arbitrary value to an enum typed variable, where the value does not need to be in the domain of the enum. I.e., the code
\begin{verbatim}
enum Color { BLACK = 0, WHITE = 1} col;
col = (Color) 42;
\end{verbatim}
is completely legal in C++ (dating back to the C ancestry) and will not raise any warning at compile or run time. In Java the cast is illegal and the code will be rejected by the compiler. The company decided to clean up the use of their database interface, where at many call sites an integer constant was passed for an enum, sometimes with unsupported values, mixed up argument order and a return value that did not fit the databases vendors specification. In C++ all of this was legal by the language rules although semantically incorrect. The Java type system is used here to detect these defects and correct the C++ so that the transpiler emitted correct Java code.

The last component of the transpiler not yet described is the concept rewriter in the lower left corner of figure~\ref{fig:intstruct}. This simple box in fact represents a collection of classes that interpret several nodes of the abstract syntax tree at once. The result of the interpretation ranges from omitting code generation for specific C++ classes over emitting it unchanged (the default) to completely rewriting the class. In the next section several of the rewriters will be explained in more detail.

\section{Addressing the Challenges}\label{sec:addressingTheChallenges}

Converting one programming language into another always poses challenges, because every language has its own features, design decisions, constraints and deficiencies. Although C++ and Java are not as far apart as say C++ and a relational language like Prolog~\cite{prolog} but when looking at the design principles are the differences enough to cause some serious headache.

\subsection{Multiple inheritance}

As already reported had the C++ code base some hundred classes that inherited from multiple base classes. Java allows inheritance from one base class only, but supports implementation of several interfaces. An interface on the other hand can not have any members and starting from Java version 8 only default implementations for methods. Without the ability to access members those methods have very limited use, because when they are to work on data, the data needs to be passed as argument. Workarounds with defining classes within the interface are feasible, but are no replacement for multiple inheritance. Therefore some other solution had to be tailored to convert the C++ to usable Java code, that still remains read- and understandable by a developer for future maintenance.

\begin{figure}[btp]
\centering
\includegraphics[scale=0.7]{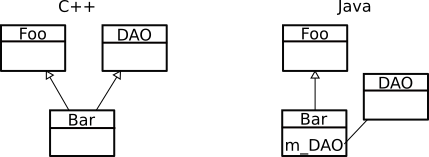}
\caption{Convert ``is a'' to ``has a'' relationship for DAOs}
\label{fig:isa2hasaRelConv}
\end{figure}

Above it was already disclosed, that in some $220$ cases one of the base classes that are part of multiple inheritance is a DAO. In those cases it is decided to convert the ``is a'' relationship into a ``has a'' relationship, like it is depicted in figure~\ref{fig:isa2hasaRelConv}. On the left hand side of the figure a multiple inheritance is shown where the class \cxxcode{Bar} inherits from \cxxcode{Foo} and a \cxxcode{DAO} class. On the right hand side the figure shows how this is restructured. The inheritance of \javacode{Foo} by \javacode{Bar} is preserved while the DAO is made a member of the \javacode{Bar} class. Of course this meant to rewrite all data accesses of the DAO's members from \cxxcode{this.getDAOMember()} to \javacode{m\_oDAO.getDAOMember()} (likewise for the setters). Here the C++ type system is sufficient to identify DAO member accesses, because all DAOs had a common base class provided by the database vendor, which allowed to identify the base type of the member access on the C++ side. This could also be used, when the data access was higher up in the inheritance hierarchy, e.g., when a class \cxxcode{Elaborate} inherited from a class \cxxcode{Base} which in turn inherited the DAO.

\begin{figure}[btp]
\centering
\includegraphics[scale=0.9]{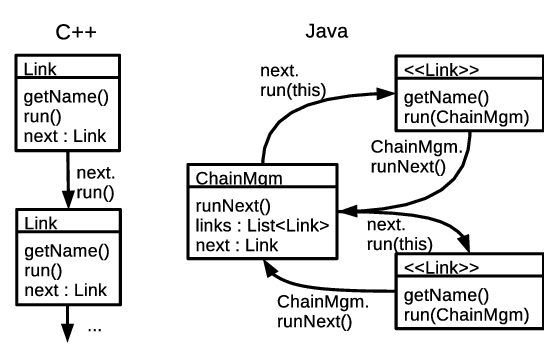}
\caption{Chain of Command in C++ and Java}
\label{fig:cocCXXJava}
\end{figure}

For the case where one of the two inherited classes is a Chain of Command class, a different solution is applied. The Chain of Command pattern as it is used in the artifact is configurable and not fixed in the code. The links of the Chain are pieced together from a control file. Each link implements a specific simple interface. The interface mandates two methods to implement: i) get the link's name and ii) run the chain. Figure~\ref{fig:cocCXXJava} conceptually shows the processing in C++ and its transformation to Java. On the left hand side the C++ traversal of the Chain is just following the \cxxcode{next} pointers. While on the right hand side, the Java implementation alternates control of flow between a chain manager and the links. The chain manager calls the \javacode{run} method of the next link supplying itself as argument to enable the link to hand back control by calling the chain manager's \javacode{runNext} method. The chain manager's \javacode{runNext} method bumps the \javacode{next} pointer along the list of \javacode{links} and calls the \javacode{run} method on the next link. When the list of links ends or a link decides to stop execution of the chain, e.g. because of an error, then the Java stack takes care of traversing the chain backwards. On start up of the software every link class is registered at a chain builder object. This is coded explicitly and for this being able to compose the chains getting the link's name is required. The chain builder is responsible for setting up the chains from the configuration.

In the case where one class inherited from is a DAO and the other the Chain class, both inheritances are removed and the above approaches are combined. I.e., the DAO's relationship to the class is transformed into as ``has a'' and the class implements the Link interface. Note that the Chain transformation is totally mechanical. The \cxxcode{getName} method is implemented in the inheriting class in C++ already and the \cxxcode{run} method of the Chain base class, is calling \cxxcode{run} on a private \cxxcode{next} pointer, if that is set. This forces classes inheriting from the Chain class to call the base classes \cxxcode{run} method to traverse the Chain. In Java the interface of the \javacode{run} method is extended by the chain manager object and the call of the base classes \cxxcode{run} method is rewritten to \javacode{ChainMgm.runNext()}.

Classes with multiple inheritance that do not fit in the above categories are investigated further. In most cases the C++ code can be re-factored to model one or more base classes as Java compatible interfaces. In C++ these base classes were transformed into abstract classes without members and only either pure virtual functions or default implementations. One of the concept rewriter's components analyzed whole classes to learn whether they resemble interfaces. In Java the classes inheritance of the interface resembling class is then rewritten to be an implements.

Unfortunately is this technique not applicable to all classes with multiple inheritance. A few (less than five) classes had more than one base class that were not transformable in the described ways. These few classes were converted to Java manually and a guard added to the build process, so that changes in C++ to the classes or its base classes produced a notification to check that these classes are still correct. This was the most feasible and cost effective way to handle these few cases.

\subsection{Stream-based in- and output}

C++'s stream based input and output facility can be easily transformed into a concatenation of strings into a \javacode{StringBuilder}. The content of the concatenation is then either printed on the console or emitted to a file using Java's intrinsic facilities or stored in a string for \cxxcode{stringstreams}. The transpiler visitor analyzed the type of the left hand sided argument of both shift operators to deduce, whether stream IO is at hand. If stream IO is detected then a property in the visitor is set to give the type of stream, i.e., console output or input or stringstream or general in- or out-stream. This differentiation is only necessary to emit neat Java and keep it readable. Knowing that a standard output stream is used in C++, a terminating \cxxcode{std::endl} functor is converted to Java's \javacode{System.out.println(...)}, while without the \cxxcode{std::endl} just a \javacode{System.out.print(...)} is emitted. This sounds like a huge effort for something rather seldom used, but keep in mind, that the C++ is to be thrown away and the Java code needs to be maintainable by developers.

In Figure~\ref{fig:streamio} some very simple C++-stream based output is shown in the upper part. The resulting output is expected to be \verb|Hello,42| and \verb|Output w/o an endl|, each on a separate line ending in a newline. The lower part shows the emitted Java. The \javacode{StringBuilder} is chosen for string concatenation, because it is efficient and able to convert most of the data to emit, i.e., it is able to convert an int to a string and so on. The last line in the figure shows, that where no concatenation is required the transpiler visitor detects this and omits the creation of an unneeded \javacode{StringBuilder} printing the string immediately, but using the \javacode{System.out.print} method, because no \cxxcode{std::endl} is on the input.

\begin{figure}[btp]
C++
\begin{verbatim}
std::cout << "Hello" << ',' << 42
    << std::endl
std::cout << "Output w/o an endl\n";
\end{verbatim}
Java:
\begin{verbatim}
System.out.println(new StringBuilder()
   .append("Hello").append(',')
   .append(42));
System.out.print("Output w/o an endl\n");
\end{verbatim}
\caption{Translating C++ stream IO to Java's IO system}
\label{fig:streamio}
\end{figure}

\subsection{enum and integer conversion}

The ancestry of C++ from C is still visible among other things in how C++ handles enums. An enum in C++-98 is some numeric data type with some predefined constants. But the value of a variable typed as an enum can have any value that fits into the underlying numeric data type, usually an int.

Java on the contrary allows an enum to only have the constant values declared. Furthermore, enums in Java are not assignable to int typed variables and casting an int to an enum is a syntax error. Unfortunately is this used a lot of times in the artifact to convert. Fortunately, the most enums are used with the database interface. The two most used enums are the return code of an operation and the other flags whether empty results are allowed or not.

Enabling the Java code to behave and look alike the C++, each enum is converted to a more elaborate structure. Figure~\ref{fig:javaenum} gives an example on the simple C++ \cxxcode{enum Color \{RED = 0, BLACK, WHITE = 3\}}. I.e. the enum is treated like a class, although the enum keyword is used for its declaration. A constructor taking an int as argument is generated which stores the int value (without checking the values validity) in the private member that is also generated at the end of the structure. The enum values given in C++ are converted to calls of the constructor supplying the value used or computed from the C++ code. In the example no value for \cxxcode{BLACK} is given in C++, which therefore computes by the C++ rules to the highest used value in the enum so far plus one. Now referencing an enum value in C++, say \cxxcode{WHITE}, will be translated to Java as \javacode{Color.WHITE}, which references the identifier \javacode{WHITE} in the enum \javacode{Color} which in Java is an object. I.e., the Java representation of enums is not as lightweight as in C++.

\begin{figure}[btp]
\begin{verbatim}
public enum Color {
  RED(0), BLACK(1), WHITE(3);

  Color(int val) {
    m_Value = val;
  }

  public int asNum() {
    return m_Value;
  }

  public static Color forNum(int val) {
    switch (val) {
      case 0: return RED;
      case 1: return BLACK;
      case 3: return WHITE;
      default:
        throw new RuntimeException(
          "Invalid enum value");
    }
  }

  private int m_Value;
}
\end{verbatim}
\caption{Java representation of the C++ \cxxcode{enum Color \{RED = 0, BLACK, WHITE = 3\};}}
\label{fig:javaenum}
\end{figure}

Figure~\ref{fig:javaenum} also shows that two methods are generated. \javacode{Color::asNum()} allows to retrieve the numerical value of the enum. It is employed where in C++ the integer representation of the enum is desired. Optimization has been done when generating Java to prevent converting to integer on both sides of an in-/equality comparison just to make the Java look more neat. The second method works the other way around. For an integer \javacode{val} the enum-object is returned or when the value is not representable in an enum value then an exception is thrown. Note, that this routine is the only way to get an enum-object. The constructor is intentionally not exposed to prevent initializing an enum-object with arbitrary values.

\subsection{Handling Constructors and Destructors}

Constructors in most cases just initialize some members of an object. Those are not a problem. Unfortunately are there also constructors in the artifact that have side effects. The most usual side effect is allocating a resource in the form of a database cursor. Database cursors in the product used are expensive in that they are limited and require memory in the DBMS. Therefore managing them efficiently is mandatory. The C++ code followed the RAII pattern (resource acquisition is initialization). Whenever a class is meant to iterate over a bunch of database results the cursor is acquired as part of the object's construction. Most of the time an object of such a class is placed on the stack of a calling method. This ensured that on leaving the method in any way possible (by return or exception), the object is destroyed and in this way the resource freed. Unfortunately does Java not have destructors. The language uses a garbage collector that from time to time expires unused objects, but does not call some kind of destructor on it. The memory of the object is just returned to the pool of free memory.

This deficiency has also been recognized by the Java language developers. From Java version 7 on, the paradigm of \textit{try with resources} is introduced. Any object that implements the \javacode{java.io.Closeable} interface is able to free allocated resources in its \javacode{close} method. One or more of these objects can be created in the ``head''-line of a try-statement as depicted in figure~\ref{fig:trywithres}. Here resources \javacode{res1} and \javacode{res2} are created, can be used in the block of the try and will be closed when the try-block is left. It does not matter what the reason for leaving the try-block is (end of block, return, or exception), the close method is always executed when the block is left before any following code is executed.

\begin{figure}[btp]
\begin{verbatim}
try (Resource1 res1 = new Resource1();
  Resource2 res2 = new Resource2() ) {

  // use resources

} // res1.close() and res2.close()
  // will be called here.
\end{verbatim}
\caption{Try with resources}
\label{fig:trywithres}
\end{figure}

This paradigm could now be employed whenever a destructor in C++ was declared. In the given artifact this would have lead to a lot of unnecessary try with resources inserted, because many destructors declared are empty. To prevent any further in depth analysis -- as interesting as it might have been -- it was decided to just annotate the C++ classes by a comment to indicate, that they made use of resources and required the try with resources paradigm to be employed. Of course is this prone to error. Given the small number of occurrences necessitating the annotation, management decided to spent only very limited time on this.

\subsection{Treating exceptions}

In the C++ artifact throwing exceptions and their handling is not present. Where third party libraries used them, they were encapsulated in a wrapper and an error condition returned using error codes. In some rare cases exception handling of third party libraries is ignored, because these error condition are fatal in any case and the application had to be terminated. A keep alive script furthermore is reporting an application crash, including those of not handling exceptions thrown, and restarts the application with nearly no downtime.

In the Java world this is somewhat different though. First of all are exceptions more often used by third party products to signal exceptional states. Second does restarting the Java application take significantly longer due to loading the different packages doing dependency injection and so on. Unfortunately at the time the author left the project the decision of how to handle exceptions more thoroughly had not been made. In discussion are the approaches to ignore them, add handling to the chain manager when starting to process the whole chain or at each link, i.e., add a try-catch around each invocation of the link's \javacode{run} method. The most favored was the last one having the drawback of potentially be costly.

\section{Conclusion and Results}\label{sec:conclusionAndResults}

\begin{figure}[btp]
\centering
\includegraphics[scale=0.4]{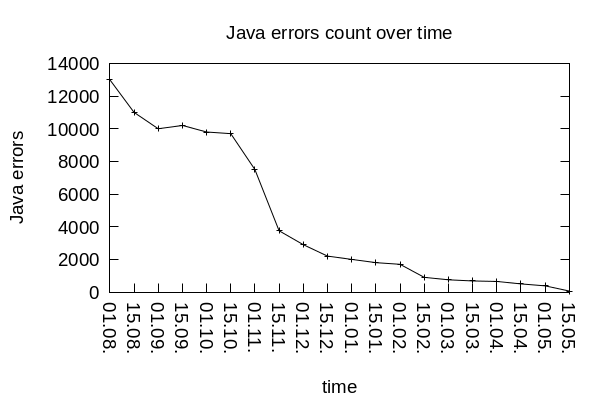}
\caption{Decrease of the Java errors over time}
\label{fig:errorovertime}
\end{figure}

The necessity to migrate an existing large C++ artifact to Java while at the same time changing the architecture and ongoing development of the C++ code base mandated a tool driven conversion methodology. A clang-tool based approach allowed a repeatable and iterative conversion of the existing code base. Figure~\ref{fig:errorovertime} gives the decrease of the Java compilation errors over the time of development. The steepest decrease in error count from 15.10. to 15.11. is motivated by the introduction of the Java type system allowing to correctly cast types and identify enum to integer and vice versa conversion mistakes.

Another learning is that at the end of the project the time spend to fix a single error increases significantly. The author left the project, when about $60$ errors remained. Those were manually corrected and the resulting Java code executed fine and some simple operations were possible rendering the whole effort a big success.

\bibliographystyle{IEEEtran}
\bibliography{CXX-Java-transpiler,isostd}

\begin{thebibliography}{10}
\providecommand{\url}[1]{#1}
\csname url@samestyle\endcsname
\providecommand{\newblock}{\relax}
\providecommand{\bibinfo}[2]{#2}
\providecommand{\BIBentrySTDinterwordspacing}{\spaceskip=0pt\relax}
\providecommand{\BIBentryALTinterwordstretchfactor}{4}
\providecommand{\BIBentryALTinterwordspacing}{\spaceskip=\fontdimen2\font plus
\BIBentryALTinterwordstretchfactor\fontdimen3\font minus
  \fontdimen4\font\relax}
\providecommand{\BIBforeignlanguage}[2]{{%
\expandafter\ifx\csname l@#1\endcsname\relax
\typeout{** WARNING: IEEEtran.bst: No hyphenation pattern has been}%
\typeout{** loaded for the language `#1'. Using the pattern for}%
\typeout{** the default language instead.}%
\else
\language=\csname l@#1\endcsname
\fi
#2}}
\providecommand{\BIBdecl}{\relax}
\BIBdecl

\bibitem{LLVM:CGO04}
C.~{Lattner} and V.~{Adve}, ``Llvm: a compilation framework for lifelong
  program analysis transformation,'' in \emph{International Symposium on Code
  Generation and Optimization, 2004. CGO 2004.}, 2004, pp. 75--86.

\bibitem{ISO:1998:IIP}
{ISO}, \emph{{ISO\slash IEC 14882:1998}: {Programming} languages ---
  {C++}}.\hskip 1em plus 0.5em minus 0.4em\relax Geneva, Switzerland:
  International Organization for Standardization, Sep. 1998.

\bibitem{Java-8}
J.~Gosling, B.~Joy, G.~L. Steele, G.~Bracha, and A.~Buckley, \emph{The Java
  Language Specification, Java SE 8 Edition}, 1st~ed.\hskip 1em plus 0.5em
  minus 0.4em\relax Addison-Wesley Professional, 2014.

\bibitem{mechturk}
\BIBentryALTinterwordspacing
D.~Difallah, E.~Filatova, and P.~Ipeirotis, ``Demographics and dynamics of
  mechanical turk workers,'' in \emph{Proceedings of the Eleventh ACM
  International Conference on Web Search and Data Mining}, ser. WSDM '18.\hskip
  1em plus 0.5em minus 0.4em\relax New York, NY, USA: Association for Computing
  Machinery, 2018, p. 135–143. [Online]. Available:
  \url{https://doi.org/10.1145/3159652.3159661}
\BIBentrySTDinterwordspacing

\bibitem{transpiler}
\BIBentryALTinterwordspacing
Devopedia. (2019) Transpiler. [Online]. Available:
  \url{https://devopedia.org/transpiler}
\BIBentrySTDinterwordspacing

\bibitem{tang:cpp2javaconv}
\BIBentryALTinterwordspacing
``C++ to java converter.'' [Online]. Available:
  \url{https://www.tangiblesoftwaresolutions.com/product\_details/cplusplus\_to\_
  java\_converter\_details.html}
\BIBentrySTDinterwordspacing

\bibitem{mohca}
S.~{Malabarba}, P.~{Devanbu}, and A.~{Stearns}, ``Mohca-java: a tool for c++ to
  java conversion support,'' in \emph{Proceedings of the 1999 International
  Conference on Software Engineering (IEEE Cat. No.99CB37002)}, 1999, pp.
  650--653.

\bibitem{barcode}
R.~C. Palmer, \emph{The Bar Code Book}.\hskip 1em plus 0.5em minus 0.4em\relax
  Trafford Publishing, 2007.

\bibitem{stl}
P.~Plauger, M.~Lee, D.~Musser, and A.~A. Stepanov, \emph{C++ Standard Template
  Library}, 1st~ed.\hskip 1em plus 0.5em minus 0.4em\relax USA: Prentice Hall
  PTR, 2000.

\bibitem{sockets}
D.~Makofske, M.~Donahoo, and K.~Calvert, \emph{TCP/IP Sockets in C: Practical
  Guide for Programmers}, 04 2004.

\bibitem{http11}
R.~Fielding, J.~Gettys, J.~Mogul, H.~Nielsen, L.~Masinter, P.~Leach, and
  T.~Berners-Lee, ``Hypertext transfer protocol -- http/1.1,'' 01 1999.

\bibitem{mpi}
L.~Clarke, I.~Glendinning, and R.~Hempel, ``The mpi message passing interface
  standard,'' in \emph{Programming Environments for Massively Parallel
  Distributed Systems}, K.~M. Decker and R.~M. Rehmann, Eds.\hskip 1em plus
  0.5em minus 0.4em\relax Basel: Birkh{\"a}user Basel, 1994, pp. 213--218.

\bibitem{dao}
H.~Feddema and R.~Petrusha, \emph{Access \& DAO Object Models: The Definitive
  Reference}, 1st~ed.\hskip 1em plus 0.5em minus 0.4em\relax USA: O'Reilly \&
  Associates, Inc., 1999.

\bibitem{bryla_loney_2007}
B.~Bryla and K.~Loney, \emph{Oracle Database 11g DBA Handbook}.\hskip 1em plus
  0.5em minus 0.4em\relax McGraw Hill, Dec 2007.

\bibitem{design_patterns}
E.~Gamma, R.~Helm, R.~Johnson, and J.~Vlissides, \emph{Design Patterns:
  Elements of Reusable Object-Oriented Software}.\hskip 1em plus 0.5em minus
  0.4em\relax USA: Addison-Wesley Longman Publishing Co., Inc., 1995.

\bibitem{prolog}
M.~Bramer, \emph{Logic Programming with Prolog}, 2nd~ed.\hskip 1em plus 0.5em
  minus 0.4em\relax Springer Publishing Company, Incorporated, 2014.

\end{thebibliography}

\end{document}